\documentclass[twocolumn,astrosymb]{aastex701}

\usepackage{starfont}
\usepackage{amsmath}

\begin{document}

\title{The Effect of Adiabatic Index on Radius Evolution and the Mass Loss}

\author[orcid=0009-0006-0239-7525,gname=Arick, sname=Collander]{Arick Collander}
\affiliation{Department of Mathematics, University of California, San Diego, La Jolla, CA 92093-0112, USA}
\email[show]{acollander@ucsd.edu}  

\author[orcid=0000-0002-1228-9820,gname=Eve, sname=Lee]{Eve J.~Lee} 
\affiliation{Department of Astronomy \& Astrophysics, University of California, San Diego, La Jolla, CA 92093-0424, USA}
\email[show]{evelee@ucsd.edu}

\begin{abstract}

Models that track the size evolution of exoplanets often assume a prescribed initial thermal state or a single adiabatic index to describe the planetary interior structure, the latter of which is taken to be $\gamma \sim1.4$ which is likely appropriate for evolved planets ($\gtrsim$1 Gyr). Extrapolating this high $\gamma$ to earlier ages (down to $\sim$million years old) is problematic since, according to thermodynamics, the adiabatic index of young planets is $\sim$1.2, which is low enough to drastically change how interior mass is distributed. We quantify the effect of varying the adiabatic index from 1.2 to 1.4 on the expected radius of the exoplanet over time. We find that envelopes of larger adiabatic indices start puffier with all else equal and undergo faster radius contraction with accelerated mass loss. Assumption of high $\gamma$ can therefore overestimate the effect of mass loss in shaping the exoplanetary population, especially when young planets are considered. We highlight the need for a more careful consideration of the initial thermal condition and internal mass distribution of planets in evolutionary models to properly interpret the radii measurements of exoplanets.

\end{abstract}

\section{Introduction} 
\label{sec:intro}

Radius is one of the most fundamental observable properties of exoplanets. Combined with mass measurements, the bulk densities of planets can be derived to obtain a broad sense of planet composition. However, mass measurements are possible for only a few planets, with orbital and physical properties that are amenable to radial velocity follow-up (e.g., at short enough orbital periods for a given mass; see \citealt{Zhu21}, their Figure 1 for the sensitivity limits of radial velocity) or in multi-planetary systems close to mean motion resonance to make use of transit timing variation \citep[e.g.,][]{Holman05,Agol05}. 

Fortunately, planet radius is most sensitively determined by the envelope mass fraction \citep{Lopez14}.
The large sample of radius measurements with well-characterized host stars enabled the discovery of the radius valley in the radius-period distribution separating the small rocky super-Earths from the larger sub-Neptunes with volatile content \citep{Fulton2017,Fulton2018,vaneylen2018}. This radius valley has been explained using theories of mass loss by photoevaporation \citep{Owen2013,Lopez_2013},\footnote{Photoevaporative models predicted the existence of radius valley.} core-powered mass loss \citep{Ginzburg2018,Gupta2020}, and gas accretion in late-stage gas-poor nebula \citep{Lee2021,Lee2022}, all of which generally assume sub-Neptunes to be composed of a rocky core with H/He-dominated envelope on top.

A well-known degeneracy in studies of exoplanet composition is that with only radius, mass, and orbital period, a given sub-Neptune may be explained by a low mass fraction of light volatiles (H/He) or a high mass fraction of heavier volatiles (e.g., water) \citep[e.g.,][]{Rogers2010}. In fact, an alternative hypothesis of sub-Neptunes is that they are waterworlds \citep[e.g.,][]{Zeng19,Venturini20}, supported in particular by early results of M dwarf planets that showed a tight correlation in the mass-radius space \citep{luque2022}. However, recent measurements showed that with a larger and updated dataset, such a tight correlation is no longer observed \citep[e.g.,][]{parc2024,dainese2025}. Furthermore, the radii of young planets are generally too large to be consistent with most sub-Neptunes being water-dominated \citep{rogers2025}. While there are sub-Neptunes that are confirmed to have water in their atmosphere or otherwise have high mean molecular weight \citep[e.g.,][]{Piaulet24,benneke2024}, it may be that waterworlds are the minority rather than the norm for sub-Neptunes.

In the context of the radius evolution of a H/He-dominated atmosphere, its initial thermal state is important. In particular, both the estimated size of the planet and the rate of mass loss will be sensitively determined by how the mass is distributed. In the simple picture of an adiabatic interior, its radial mass distribution is set by the adiabatic index $\gamma$ whereby $\gamma < 4/3$ would imply centrally concentrated mass profile and $\gamma > 4/3$ an outwardly concentrated mass profile. The former would be consistent with more compact atmosphere that is less affected by mass loss mechanisms that are governed by outer boundary conditions.

Models of initial gas accretion report the adiabatic index in the deep interior is low: $\gamma \sim 1.2$ \citep{Lee14,Lee15}. By contrast, mass loss and post-formation evolution calculations often simplify the initial thermal state with either an input entropy chosen to simulate the desired initial conditions (e.g., reaching a target radius \citep[e.g.,][]{Owen2016} or the contraction time to be equivalent to the disk dissipation time \citep[e.g.,][]{Tang24}) or an assumed fixed $\gamma > 4/3$ that is more consistent with molecular or atomic gas \citep[e.g.,][]{Owen17,Ginzburg2018}. \citet{Rogers24} used \texttt{MESA} to follow the gas accretion in gas-depleting nebula prior to mass loss by introducing an arbitrary internal heat to inflate a planet so that the atmospheric density at the Bondi radius matches the nebular density then turning off this heat to resolve accretion by cooling, assuming the planetary envelope is connected to the outer nebula by an isothermal layer, the latter of which governs the advection of accreting flow. 

For typical sub-Neptunes at $\sim$10--30 days, the Hill radius is the more appropriate outer boundary during gas accretion, rather than the Bondi radius. Furthermore, depending on the choice of opacity and equation of state, the outermost layer can deviate from isothermality \citep{Lee18} and can affect the nature of the outer advection \citep{Savignac24}. In general, early thermal evolution during the accretion phase is robust to these varying outer boundary conditions; however, the reason for such robustness is because the initial envelopes are centrally concentrated, governed by low $\gamma$ \citep{Lee16}, and it is not clear whether the aforementioned post-formation calculations capture the same low $\gamma$ within the planetary interior and especially at earlier epochs that transition from the initial formation. In spite of its importance, the changes in the interior $\gamma$ and its effect on the radius evolution received less attention in the literature with no explicit discussion to date.

In this paper, we seek to quantify the effects of varying the adiabatic index on the evolution of radius and mass of sub-Neptunes. In order to provide a clean representation of the underlying physics, we limit our analysis to an analytic and semi-analytic approach. This paper is structured as follows: in Section \ref{sec:Methods}, we describe how we compute the planet radius over time under passive cooling and under mass loss including photoevaporation and internal heat assisted bolometric mass loss; Section \ref{sec:Results} outlines our results; and Section \ref{sec:Discussion} summarizes our finding and explores their implications.

\section{Methods} \label{sec:Methods}

\subsection{Calculating Planet Radius Using the Envelope Mass Fraction}
We adopt the procedure outlined in \citet{Lee2021} with a few minor corrections and additions, which we summarize here. We assume the bulk of the envelope mass is locked within the inner adiabat with the outermost layers of the envelope being isothermal contributing minimally to the total mass. For a given adiabatic index $\gamma$, the density profile follows\footnote{Here, we fix the typo in equation 6 of \citet{Lee2021} which excluded the power-law index of $1/(\gamma-1)$. The rest of the procedure of \citet{Lee2021} correctly includes this power-law index, which we adopt here.}
\begin{equation}
    \rho(r) \simeq \rho_{\rm rcb}\left[\nabla_{\rm{ad}}\frac{GM_{\rm{core}}}{c_s^2}\left(\frac{1}{r}-\frac{1}{R_{\rm rcb}}\right)\right]^{\frac{1}{\gamma-1}},
\end{equation}
where $\rho_{rcb}$ is the density at the radiative-convective boundary (rcb) $R_{\rm rcb}$, $\nabla_{ad}\equiv\frac{\gamma-1}{\gamma}$ is the adiabatic gradient, $G$ is the gravitational constant, $M_{\rm core}$ is the mass of the planetary core, and $c_{\rm{s}}^2\equiv\frac{kT_{\rm{eq}}}{\mu \rm{m}_H}$ is the square of the sound speed of the planetary atmosphere. Here, $k$ is the Boltzmann constant, $T_{\rm eq}\equiv T_{\rm eff}(R_\star/a)^{1/2}$ is the equilibrium temperature of the planet, $T_{\rm{eff}}$ is the effective temperature of the star, $R_\star$ is the radius of the star, $a$ is the orbital distance of the planet, $\mu = 2.37$ is the envelope mean molecular weight for H/He-dominated atmosphere at solar metallicity, and $\rm{m}_H$ is the mass of the hydrogen atom. We can then write the total mass of the envelope as
\begin{equation}
\label{eq:2}
    M_{\rm{env}} \simeq 4\pi\rho_{\rm{rcb}}R_{\rm{rcb}}^3\left(\nabla_{\rm{ad}}\frac{GM_{\rm{core}}}{c_s^2R_{\rm{rcb}}}\right)^{\frac{1}{\gamma-1}}I_2,
\end{equation}
where $I_2$ is the structural integral of the form
\begin{equation}
\label{integral}
    I_n \equiv \int_{R_{\rm{core}}/R_{\rm{rcb}}}^1x^n(x^{-1}-1)^{\frac{1}{\gamma-1}}dx,
\end{equation}
and $R_{\rm{core}}=R_\oplus(M_{\rm{core}}/M_{\oplus})^{1/4}$ is the radius of the planetary core \citep{Valencia06}. 

By definition, at the rcb, the temperature gradient follows both radiative diffusion and that of an adiabat so we can write
\begin{equation} \label{tempgrad}
    \rho_{rcb} = \frac{64\pi\sigma_{\rm{sb}}\mu \rm{m}_H}{3k\kappa}\nabla_{ad}\frac{GM_{\rm{core}}T_{\rm{eq}}^3}{L}
\end{equation}
where $\sigma_{\rm{sb}}$ is the Stefan-Boltzmann constant, $\kappa\equiv10^C\rho^\alpha_{\rm{rcb}}(k/\mu m_H)^\alpha T_{\rm{eq}}^{\alpha+\beta}$ is the opacity at the rcb in which we use $C=-7.32$, $\alpha=0.68$, and $\beta=0.45$ \citep{Rogers2010}, and
\begin{equation} \label{Luminosity}
    L\simeq \frac{GM_{\rm{core}}M_{\rm{env}}}{\tau_{\rm{KH}}R_{\rm{rcb}}}\frac{I_1}{I_2}
\end{equation}
is the cooling luminosity where $\tau_{KH}$ is the Kelvin-Helmholtz cooling time of the envelope and $I_1$ is a structural integral again following the form given in Equation \ref{integral}. We take a range of $\tau_{kh}\in[4,1000]$ Myr; each $\tau_{kh}$ corresponds to unique $\rho_{\rm rcb}$ and $R_{\rm rcb}$ which we thread together to draw a time evolution setting $\tau_{\rm kh}$ equal to each temporal instance (see Section \ref{ssec:evolution}). We do not consider the cooling of the core for simplicity but its inclusion would generally delay the cooling and therefore contraction of the envelope that leads to slightly more enhanced mass loss but similar radius \citep[e.g.,][]{Rogers24,Tang24}.
Substituting Equation \ref{Luminosity}  into Equation \ref{tempgrad}, we obtain:
\begin{align} \label{rho2}
    \rho_{rcb}^{1+\alpha} &= \frac{64\pi\sigma_{\rm{sb}}\mu \rm{m}_H}{3k}\nabla_{\rm{ad}}10^{-C}\left(\frac{\mu \rm{m}_H}{k}\right)^\alpha \nonumber \\
    &\times T_{\rm{eq}}^{3-\alpha-\beta}\frac{I_2}{I_1}\left(\frac{\tau_{\rm{KH}}}{M_{\rm{env}}}\right)\left(\frac{R_{\rm{rcb}}}{R_{\rm{core}}}\right)R_{\rm{core}}.
\end{align}
Additionally, rearranging Equation \ref{eq:2}, we get a separate expression for $\rho_{\rm rcb}$:
\begin{align} \label{rho1}
    \rho_{\rm rcb} &= \frac{M_{\rm env}}{4\pi}\left(\frac{R_{\rm rcb}}{R_{\rm core}}\right)^{-3+1/(\gamma-1)}R_{\rm core}^{-3+1/(\gamma-1)} \nonumber
\\
    &\times \left(\nabla_{\rm ad}\frac{GM_{\rm core}}{c_s^2}\right)^{1/(1-\gamma)}I_2^{-1}.
\end{align}

We use the \texttt{root\_scalar} function from the \texttt{SciPy optimize} package to numerically solve for $R_{\rm{rcb}}/R_{\rm{core}}$ that give $\rho_{\rm{rcb}}$ satisfying both Equations \ref{rho2} and \ref{rho1}. To determine the effect of adiabatic index, we vary $\gamma\in[1.2,1.25,1.3,1.35,1.4]$. We deliberately keep $\gamma$ spatially constant to provide a clean physical picture of its effect although in reality $\gamma$ is spatially variable. In terms of mass distribution, it is the $\gamma$ in the deep interior that is relevant \citep{Lee15}---e.g., if $\gamma$ is above 4/3 even in the deep interior, then the mass is globally outwardly concentrated. We set $R_{\rm{rcb}}/R_{\rm{core}}=1$ for $M_{\rm{env}}/M_{\rm{core}}$ that give $R_{\rm{rcb}}/R_{\rm{core}}<1.05$ to save computation time, supported by the $\sim5\%$ error in {\it Kepler} transit depth measurements. 
We then make a correction for the photospheric, observable radius that is slightly larger than $R_{\rm{rcb}}$, which we obtain using the \texttt{root\_scalar} function from the \texttt{SciPy optimize} package to numerically solve for $R_{\rm{phot}}$ satisfying

\begin{equation}
    R_{\rm{phot}} = \left[\frac{1}{R_{\rm{rcb}}} + \frac{c_s^2}{GM_{\rm{pl}}}\ln\left(\frac{\rho_{\rm{ph}}}{\rho_{\rm{rcb}}}\right)\right]^{-1}
\end{equation}
where $\rho_{\rm{ph}}=(2/3)\mu m_H \rm{g}/kT_{\rm{eq}}\kappa$ is the density at the photosphere and g $=GM_{\rm{core}}/R^2_{\rm{phot}}$ is the surface gravity (see also \citealt{Tang25}). Here, we include the correction provided in \citet{Lee2022} for unrealistic values of $\rho_{\rm{rcb}} >$ 1 g cm\(^{-3}\) by taking  $\min\{\rho_{\rm{rcb}},\rho_{\rm{bottom}}\}$ where $\rho_{\rm{bottom}}$ is the approximated gas density at the bottom of the envelope given by 
\begin{equation}
    \rho_{\rm{bottom}} \sim \frac{M_{\rm{gas}}}{4\pi R_{\rm{rcb}}^2}\frac{\text{g}\mu \rm{m}_H}{kT_{\rm{eq}}}.
\end{equation}

\subsection{Mass Loss Processes}
Next, we look at the envelope mass loss of planets closest to the host star caused by hydrodynamic winds through either photoevaporation \citep{Owen2013} or internal heat \citep{Ikoma2012,Owen2016,Ginzburg2018}---we emphasize that in the latter case, the rate of mass loss is still set by the incident bolometric radiation but it is the internal heat that supplies the gas to the upper layers. While both of these processes likely operate simultaneously, \citet{Owen2024} found that there are regions of the parameter space (planet mass, orbital period) in which one process dominates over the other. Photoevaporation dominates when the penetration radius of XUV photons is below the Bondi radius because the speed of the envelope is subsonic below the Bondi radius, allowing for the propagation of XUV powered expansion throughout the entire envelope. Otherwise, bolometric mass loss dominates as the XUV photons cannot pierce beyond the supersonic region of the envelope and, therefore, cannot significantly influence expansion. The aforementioned mechanism translates to photoevaporation dominating over the majority of the observed mass-period space except for low mass ($\lesssim$1$M_\oplus$) planets with the exact size/mass of the boundary rising at higher equilibrium temperatures \citep{Owen2024,Misener2026}. Since the goal of this paper is to determine the effect of varying adiabatic indices, we isolate each of these processes in our models for better clarity. 

To start, we assume a solar-type host star for our purpose of determining the impact of the adiabatic index on both of these processes. We evolve the envelope mass over 1 Gyr according to energy limited mass loss \citep[e.g.,][]{Watson81,Lopez_2013},
\begin{equation} \label{eq:photoevaporation}
    \dot{M}_{\rm{phot}} = -\eta \frac{L_{\rm{HE}} R^3_{\rm{phot}}}{4a^2G(M_{\rm{core}}+M_{\rm{env}})},
\end{equation}
where $\eta=0.1$ is the mass-loss efficiency factor (in reality, $\eta$ is a function of the planet's gravity and incident flux, but we keep it constant for simplicity as neither factor depends on $\gamma$) and $L_{HE}$ is the high-energy luminosity (X-ray and EUV) of the star \citep{Ribas2005,Jackson2012} given by
\begin{equation}
L_{\rm{HE}}=
    \begin{cases}
        10^{-3.5}L_{\odot} & t<100 \rm{Myr}\\
        10^{-3.5}L_{\odot}\left (\frac{t}{100\rm{Myr}}\right)^{-0.86} & t\geq100 \rm{Myr}\\
     \end{cases}
\end{equation}
Here, we include a correction with a revised power law slope of 0.86 compared to that of 1.5 which the aforementioned authors used.
Such a correction accounts for the slow decay of EUV luminosity which dominates over X-ray luminosity after 100 Myrs \citep{ King2021, Karalis2025}. In reality, energy-limited approximation is often inaccurate \citep[e.g.,][]{Owen12,Owen19,Tang25} and corrections can be made to $\dot{M}_{\rm phot}$ using empirically-determined functional form encompassing various regimes of mass loss \citep[e.g.,][]{Kubyshkina18,Caldiroli22}. Our goal here is to assess the relative effect of $\gamma$ on radius evolution, and over our chosen range of $\gamma$, the initial planet radius remains within an order of magnitude such that it remains within low gravitational potential regime, ensuring the functional form of $\dot{M}_{\rm phot}$, even with the modern correction, remains consistent across $\gamma$'s. We therefore use the simpler form of $\dot{M}_{\rm phot}$ with fixed $\eta$ as it does not detract from our final conclusion.

For our internal-heat-assisted mass loss, we take the mass loss rate provided by \citet{Owen2016} which follows the classic Parker wind model:
\begin{equation} \label{eq:Internal Heat}
    \dot{M}_{\rm{ih}} = 4\pi R_{\rm{B}}^2\rho_{\rm{ph}}\exp\left(-2\frac{R_{\rm{B}}}{R_{\rm{phot}}}\right)c_s,
\end{equation}
where $R_{\rm{B}}=\frac{G(M_{\rm{env}}+M_{\rm{core}})}{2c_s^2}$ is the Bondi radius of the planet (see also \citealt{Tang24}). We do not distinguish between envelope vs.~core for the internal source of heat and uniformly adopt equation \ref{eq:Internal Heat} which applies for bolometrically driven outflows.

\section{Results} \label{sec:Results}

\begin{figure} 
\epsscale{1.15}
  \plotone{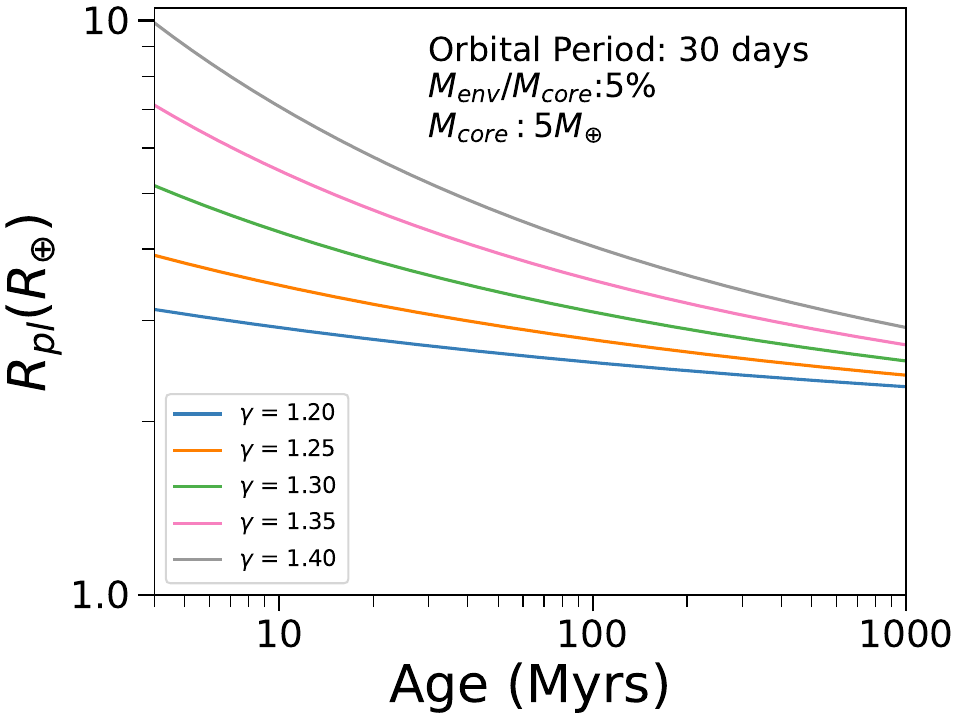}  \plotone{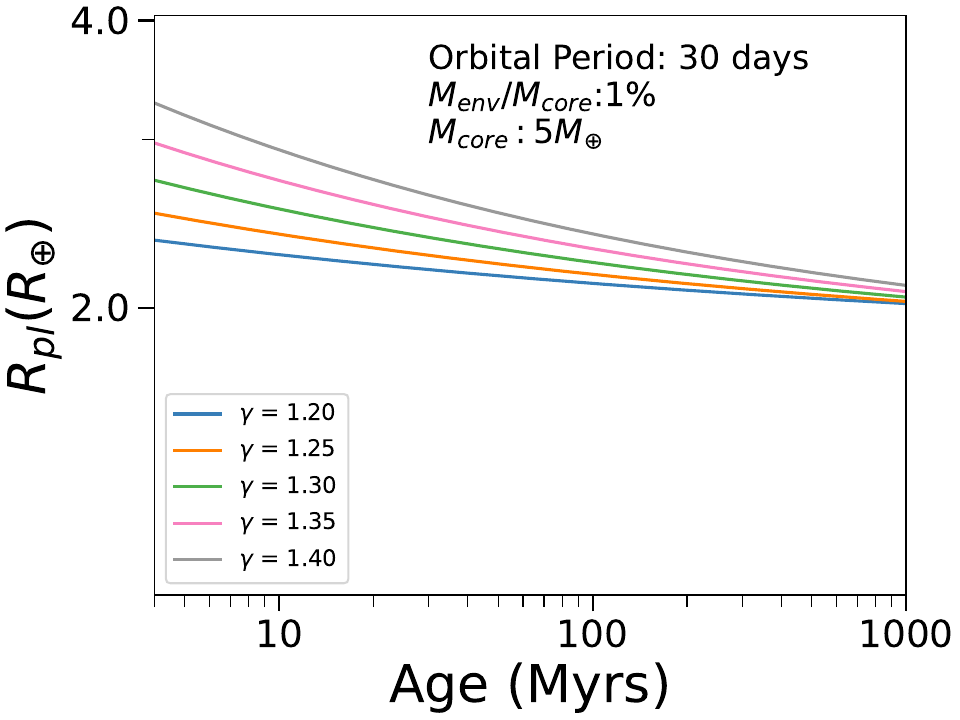}
\plotone{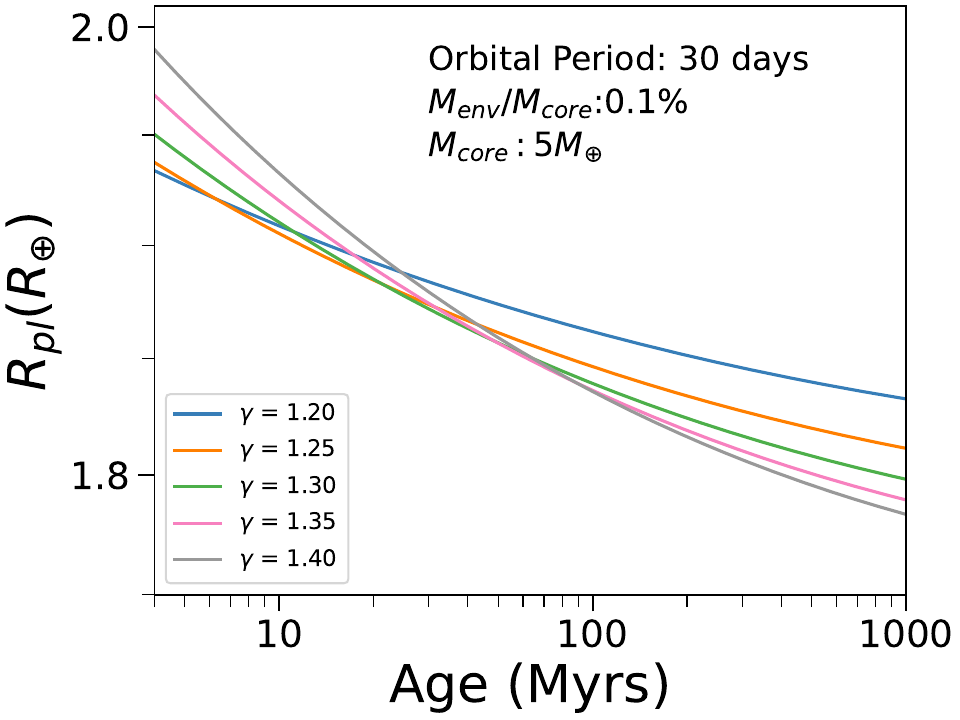}
\caption{Planet radius $R_{\rm pl}$ (set to $R_{\rm phot}$) vs planet age with $M_{\rm core}=5 M_{\oplus}$ at an orbital period of 30 days, with envelope mass fractions $M_{\rm env}/M_{\rm core}$ of 5\%, 1\%, and 0.1\% from top to bottom panels. Different colors each correspond to different $\gamma$. Larger $\gamma$ envelopes begin with larger $R_{\rm pl}$ but also contract more quickly. At the lowest $M_{\rm env}/M_{\rm core}$, the difference in the initial radius for varying $\gamma$ is small enough that the $R_{\rm pl}$ for $\gamma = 1.4$ dips below that of $\gamma = 1.2$ at $\approx25\rm{Myrs}$.}
\label{fig:radius_evo}
\end{figure}

\subsection{Planet Evolution without Mass Loss}
\label{ssec:evolution}
\begin{figure} 
\epsscale{1.15}
\plotone{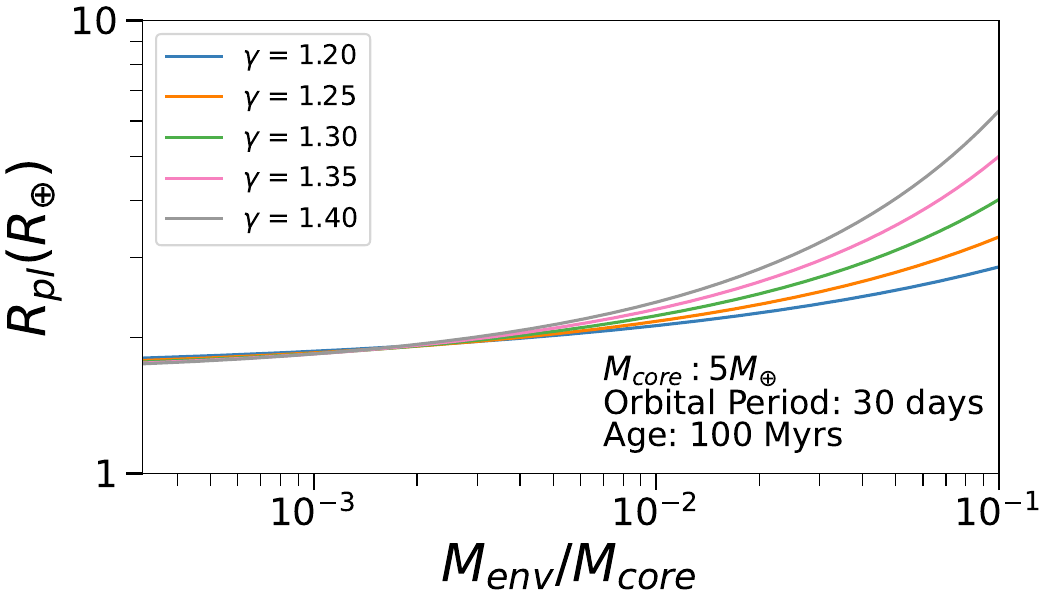}
 \caption{Planet radius $R_{\rm pl}$ vs envelope mass fraction $M_{\rm env}/M_{\rm core}$ for a planet with $M_{\rm core}=5 M_{\oplus}$, at an orbital period of 30 days, and at the age of 100 Myr. At large enough $M_{\rm env}/M_{\rm core}$, larger $\gamma$ feature larger $R_{\rm pl}$ but the radii converge to the same value at $M_{\rm env}/M_{\rm core} \sim 0.17\%$ below which lower $\gamma$ show slightly larger $R_{\rm pl}$.}
 \label{fig:Radius_env}
\end{figure}

Isolated in a vacuum, cooling planets are expected to shrink with time, which we verify in Figure \ref{fig:radius_evo} where we take $\tau_{\rm KH}$ as the planet age starting at 4 Myr. Higher $\gamma$ correspond to an outwardly concentrated mass profile, so we may expect the planet to be puffier with all else equal as compared to lower $\gamma$ at all times. This expectation bears out in our calculations but only for envelope mass fractions greater than 0.1\%. At lower envelope mass fractions, the radius of the planet begins larger for higher $\gamma$, but becomes smaller than that for lower $\gamma$ as the planet evolves (see Figure \ref{fig:Radius_env}). Such a behavior arises because high $\gamma$ envelopes shrink more rapidly in time compared to that of low $\gamma$ so that if given enough time, the two $\gamma$ envelopes reach a near-convergence in radius. In addition, for a very low mass envelope (0.1\% mass fraction), the envelope is thin enough that the difference in initial radius between $\gamma = 1.4$ and 1.2 is already small and the $\gamma=1.4$ envelope appears smaller than the $\gamma=1.2$ envelope at an evolved stage by $\lesssim$2\%. 

Heuristically, the shallower time evolution of low $\gamma$ envelopes can be understood as the planetary envelope being already compact with the mass of their gas envelope being centrally concentrated, so the change in the radius from cooling is minimal since most of the gas is already packed closely to the central core of the planet. For higher $\gamma$ envelopes, since their mass is outwardly concentrated, they are more susceptible to changes in the outer boundary and the radius contraction is more pronounced.
The dependence on $\gamma$ will naturally be more significant for thicker and heavier envelopes where there is now sufficient mass locked in the gaseous envelope that the difference in the initial radius with varying $\gamma$ is large. 

\begin{figure*}
\epsscale{1.1}
\plottwo{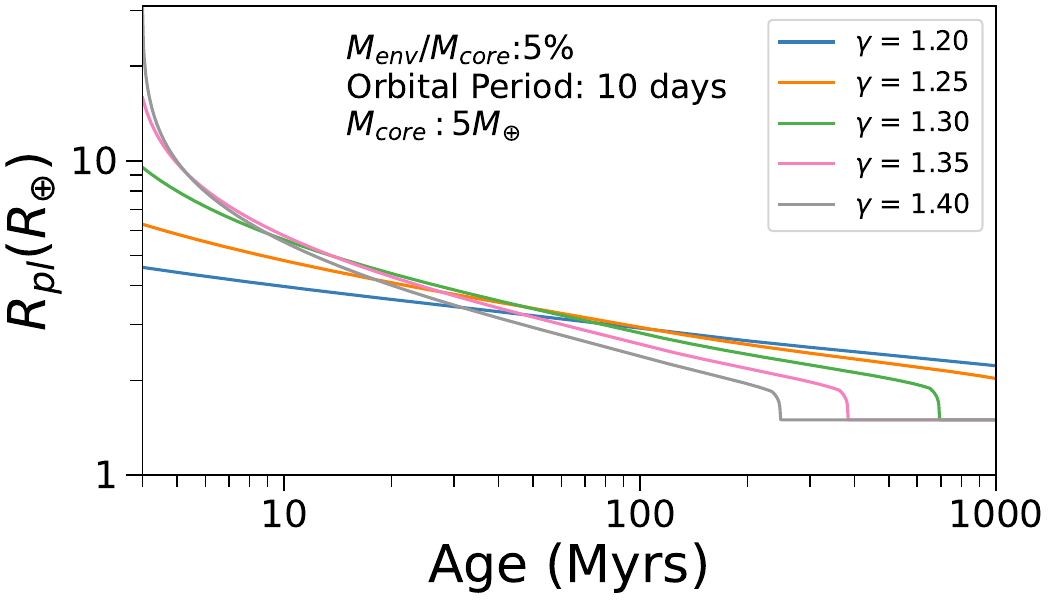}{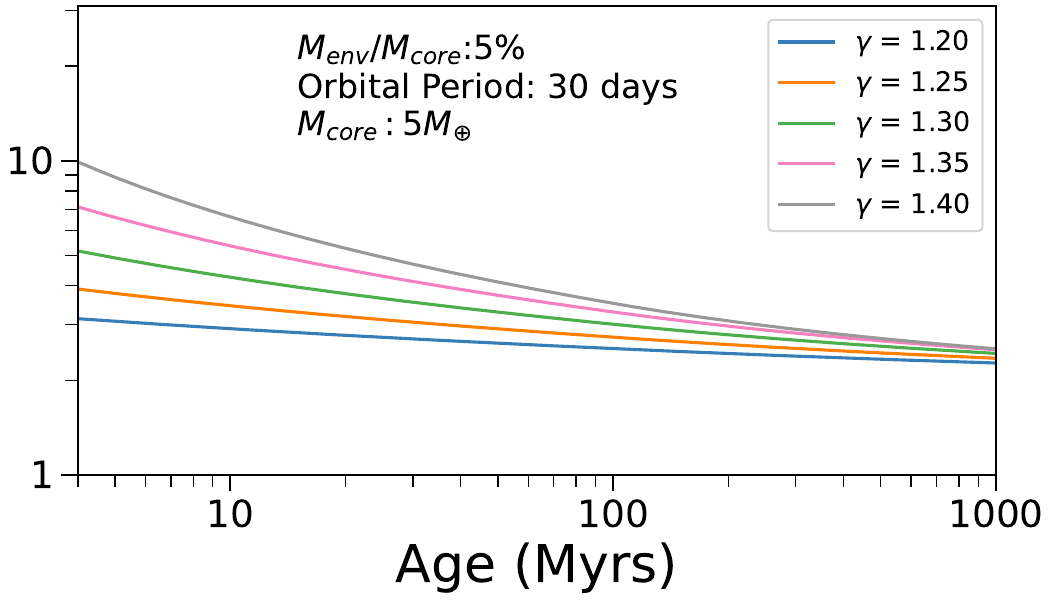}
\caption{Photoevaporation for a planet with $M_{\rm core}=5 M_{\oplus}$, $M_{\rm env}/M_{\rm core} = 5\%$, at orbital periods of 10 days (left) and 30 days (right). The rate of mass loss is higher for larger $\gamma$. At an orbital period of 10 days, $\gamma=1.4$ envelope is completely stripped away within $\sim$300 Myrs with lower $\gamma$ envelopes surviving over progressively longer timescales. At an orbital period of 30 days, all $\gamma$ approach a radius of $R_{\rm pl}\approx2.41 R_{\oplus}$.}
\label{Photoevaporation}
    \plottwo{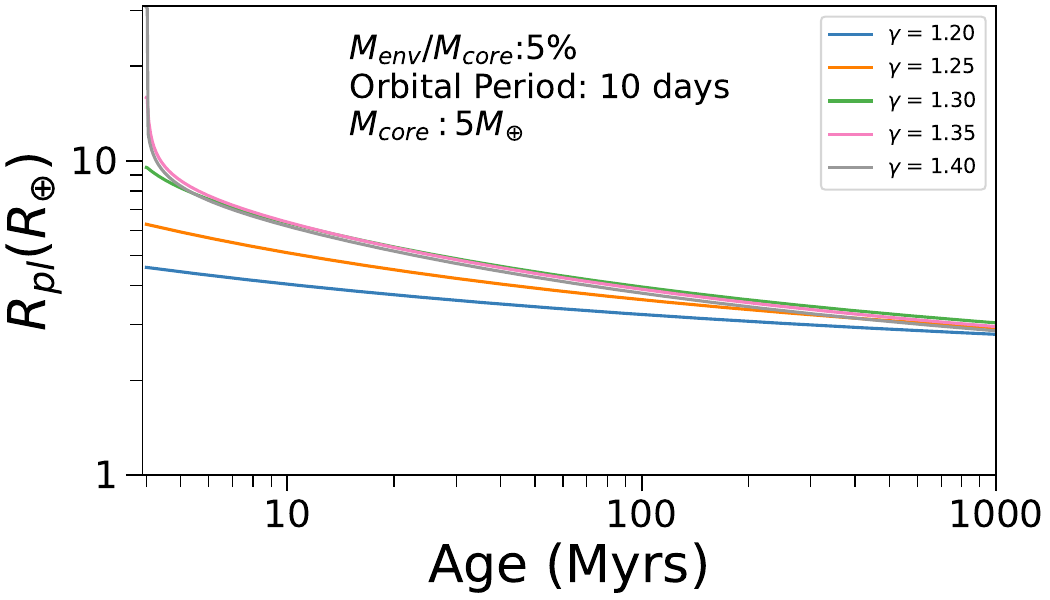}{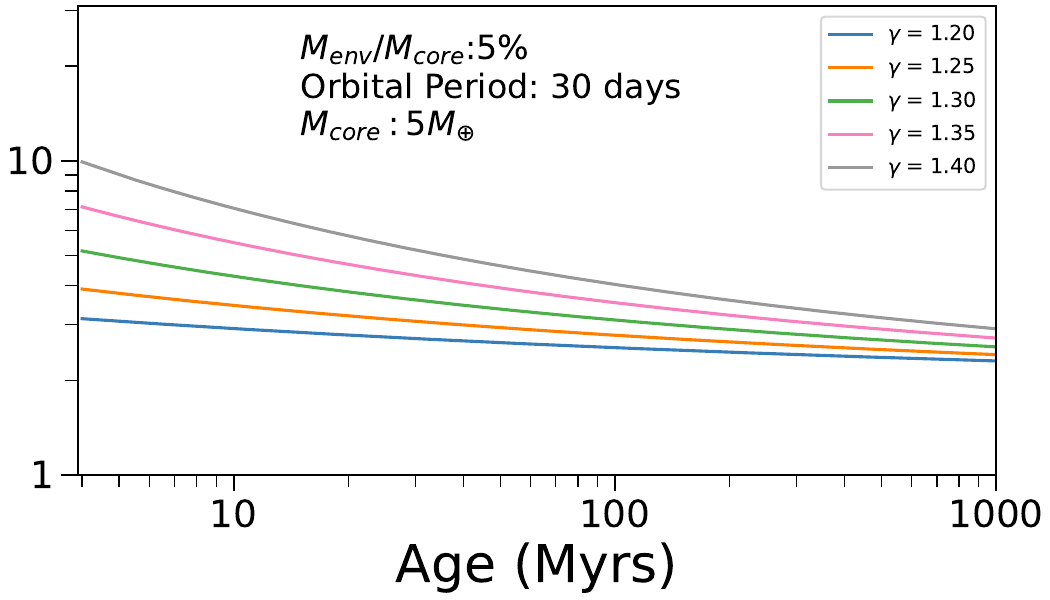}
    \caption{Bolometric mass loss for a planet with $M_{\rm core}=5 M_{\oplus}$, $M_{\rm env}/M_{\rm core} = 5\%$, at orbital periods of 10 days (left) and 30 days (right). We find radius convergence, which is reached more quickly at 10 days than 30 days.}
    \label{Internal Heat}
\end{figure*}

\subsection{Impact of $\gamma$ on Mass Loss}
\label{ssec:impact-mloss}

We now consider the effect of $\gamma$ on mass loss processes.
Larger planets where the mass is outwardly concentrated are expected to undergo more rapid mass loss as both photoevaporation and bolometric loss are dependent on the radius of the planet. Physically, in photoevaporation, a larger planet will not only be prone to having more of its envelope experiencing less of a gravitation pull by the planet, but will also have a larger surface area with which to absorb XUV radiation from the host star. In the case of  bolometric loss, an outward mass concentration implies that more of the envelope mass will be concentrated near the Bondi radius and will be more readily stripped away as the envelope expands. We verify these radius relations mathematically from Equations \ref{eq:photoevaporation} and \ref{eq:Internal Heat}. 

Since high $\gamma$ envelopes start with larger radii, we expect they will correspond to more rapid mass loss, which we verify in Figures \ref{Photoevaporation} and \ref{Internal Heat}. 
Under photoevaporation at an orbital period of 10 days, we obtain much higher mass loss rates for higher $\gamma$, leading to none, except for $\gamma = 1.2$ and 1.25, surviving beyond 1 Gyr. For an orbital period of 30 days, the radius convergence that was observed in Figure \ref{fig:radius_evo} is magnified because the planetary radius for a higher adiabatic index decreases more rapidly as the increased mass loss accelerates the shrinkage in radius.

For bolometric mass loss, we observe a similar pattern as illustrated in Figure \ref{Internal Heat} but with more subdued level of mass loss concluding with radius convergence between all $\gamma$'s.

\begin{figure*}
\gridline{\fig{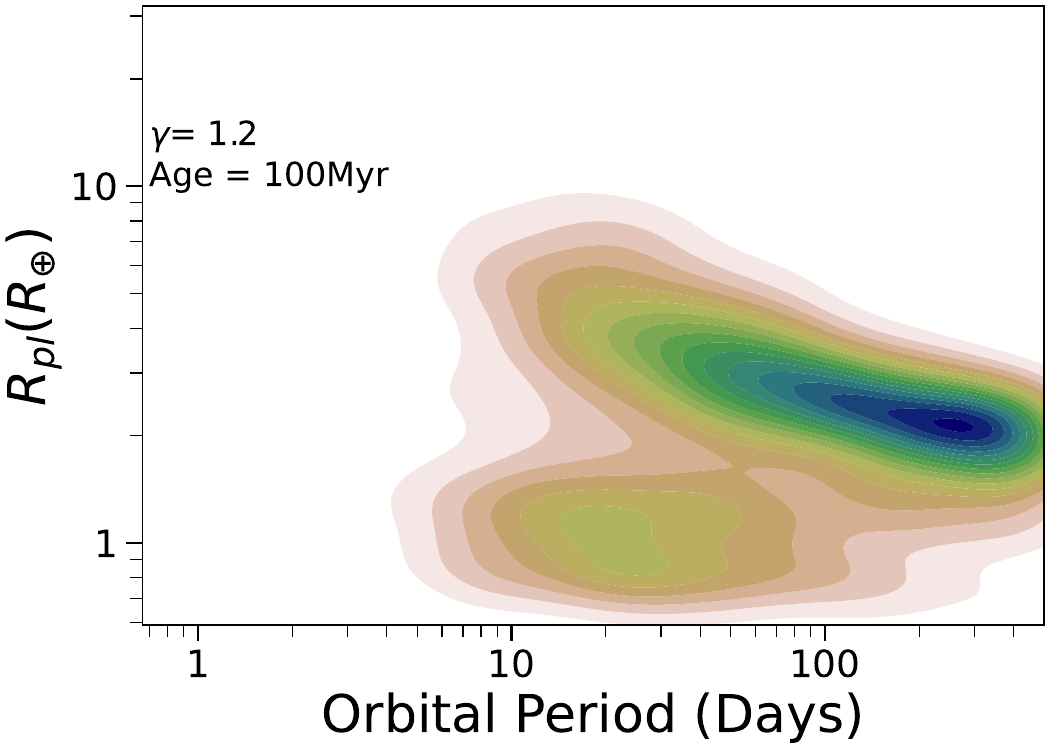}{0.5\textwidth}{}
                     \fig{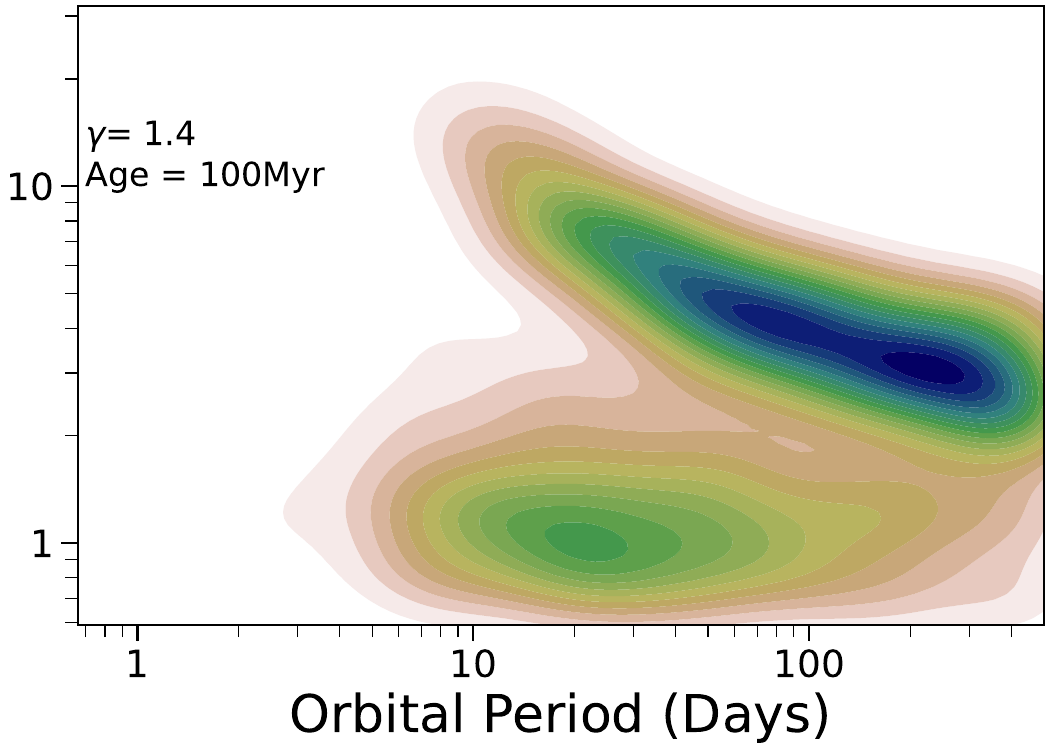}{0.5 \textwidth}{}}
\gridline{\fig{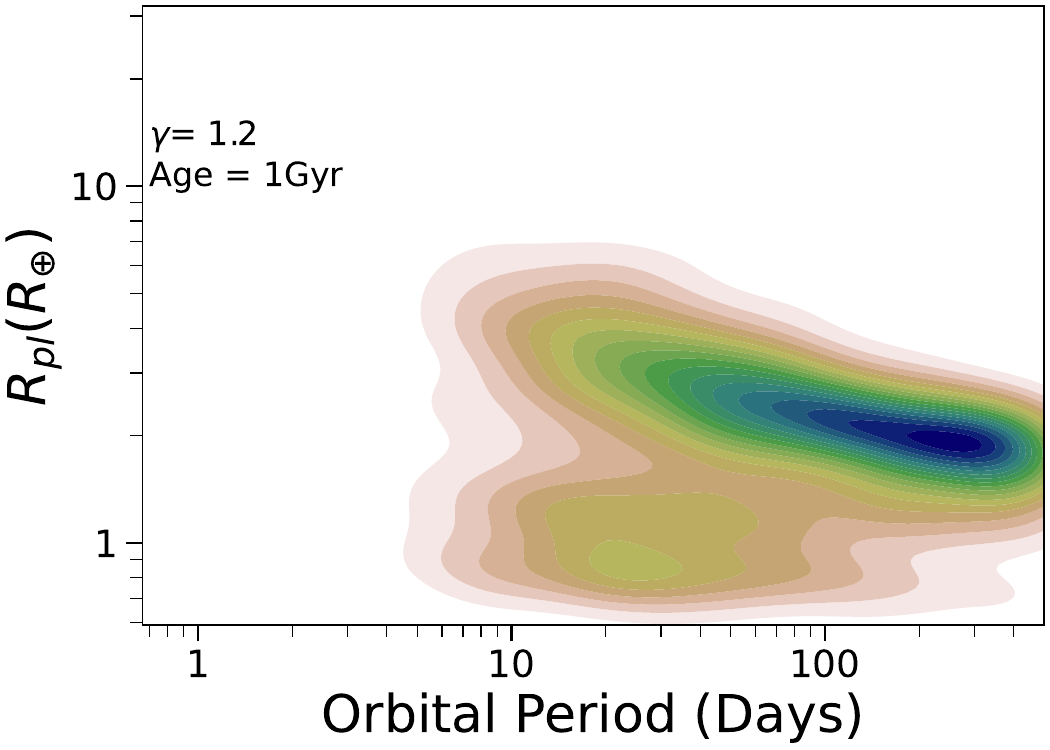}{0.5\textwidth}{}
                    \fig{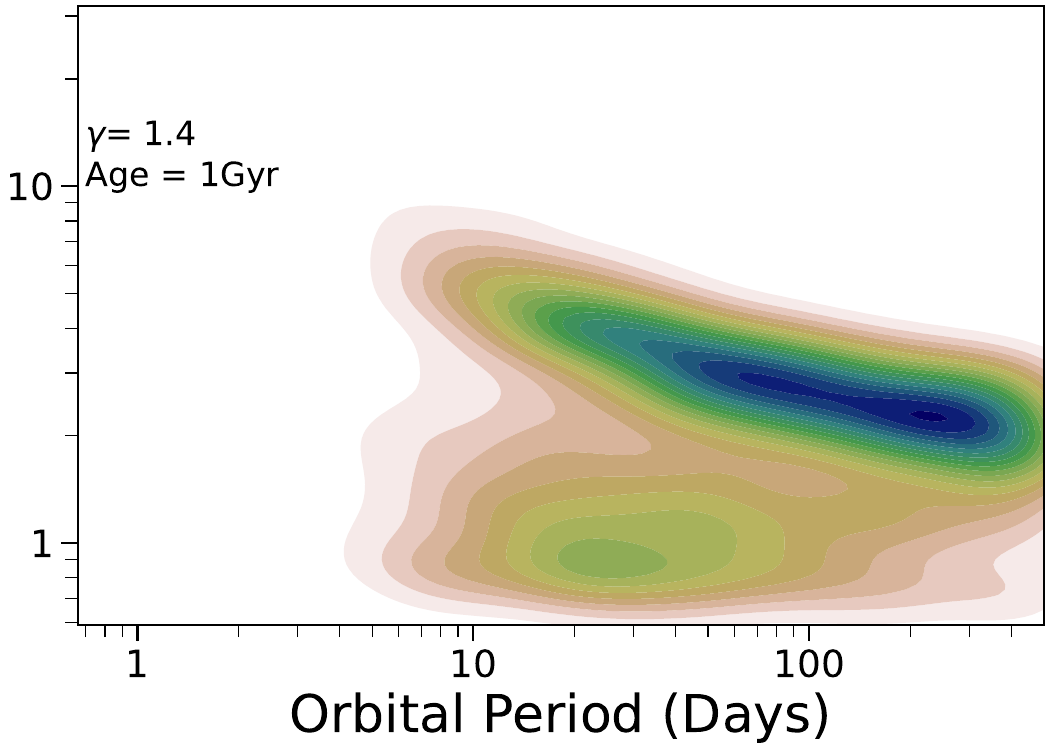}{0.5\textwidth}{}}
\caption{Planet radius $R_{\rm pl}$ vs orbital period without mass loss using model planets constructed following \citet{Lee2022} (see the main text for more details) illustrated using Gaussian kernel density estimation with Scott's rule for bandwidth selection. The top row corresponds to a planet age of 100Myr and the bottom row corresponds to a planet age of 1Gyr. The left column corresponds to $\gamma=1.2$ and the right column corresponds to $\gamma=1.4$. We see more pronounced population of super-Earths and the radius valley appears more filled in for $\gamma = 1.4$ compared to $\gamma = 1.2$.}
\label{fig:Radius_kde_5800K}
\end{figure*}
\begin{figure*}
    \gridline{\fig{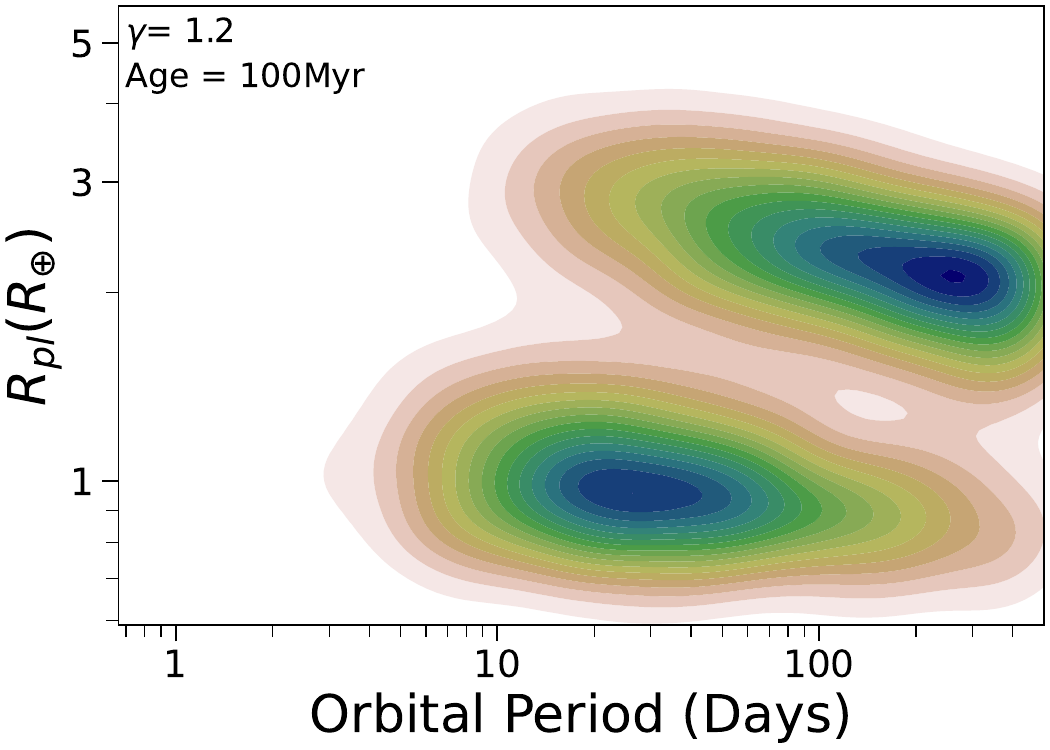}{0.5\textwidth}{}
                     \fig{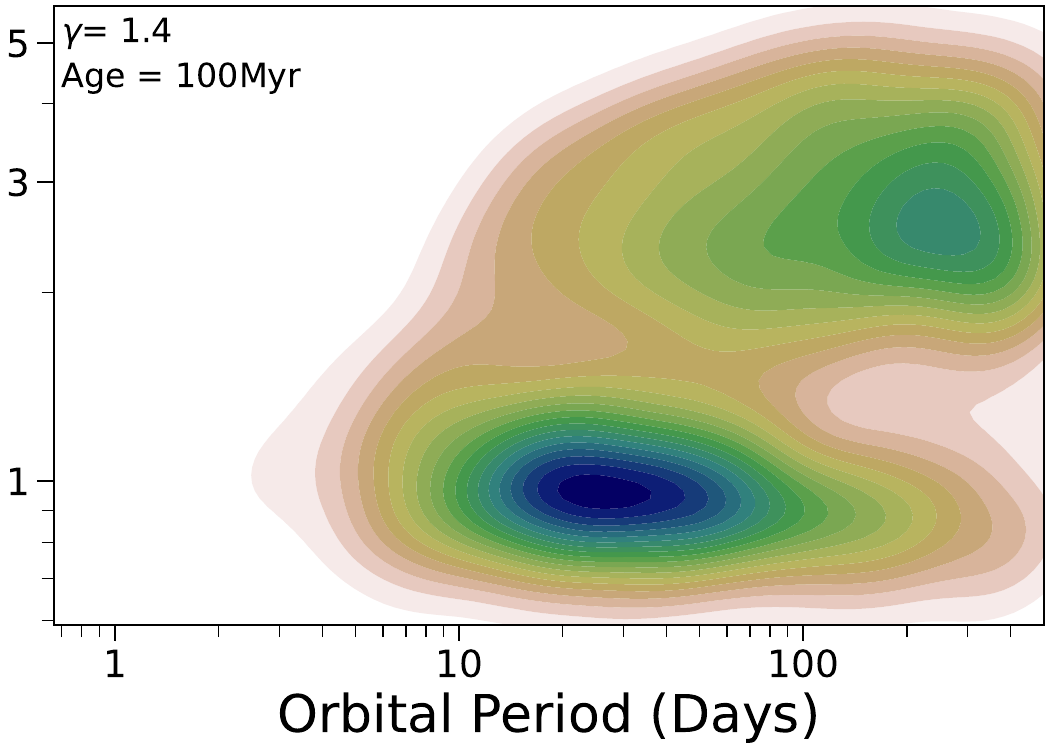}{0.5 \textwidth}{}}
\gridline{\fig{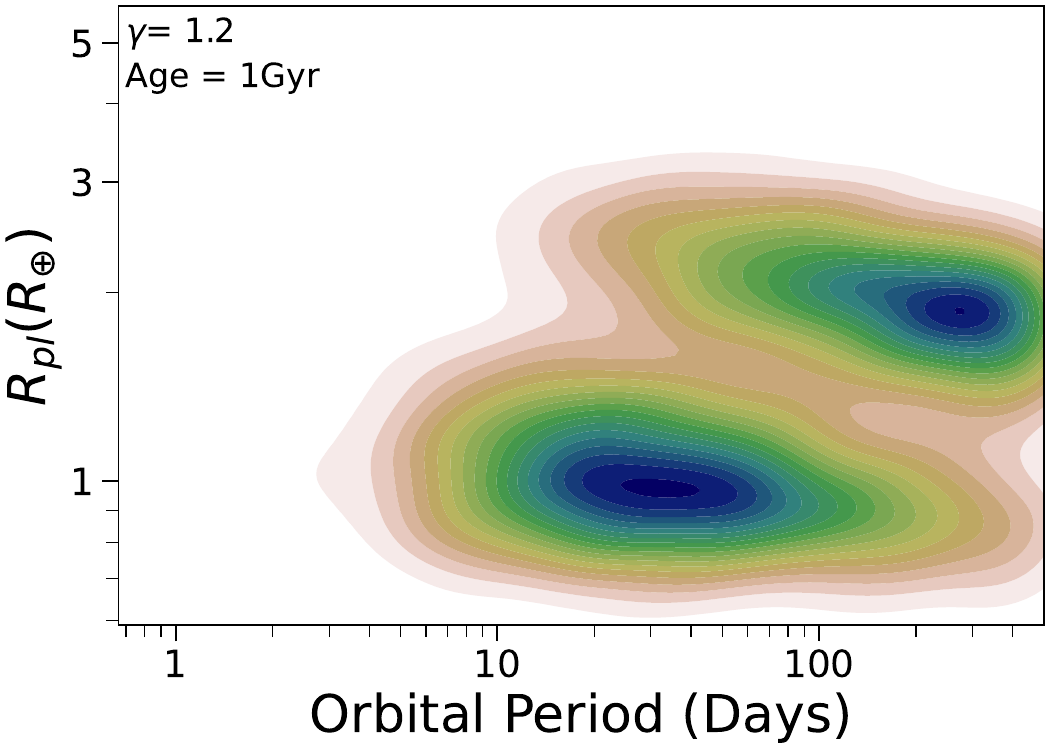}{0.5\textwidth}{}
                    \fig{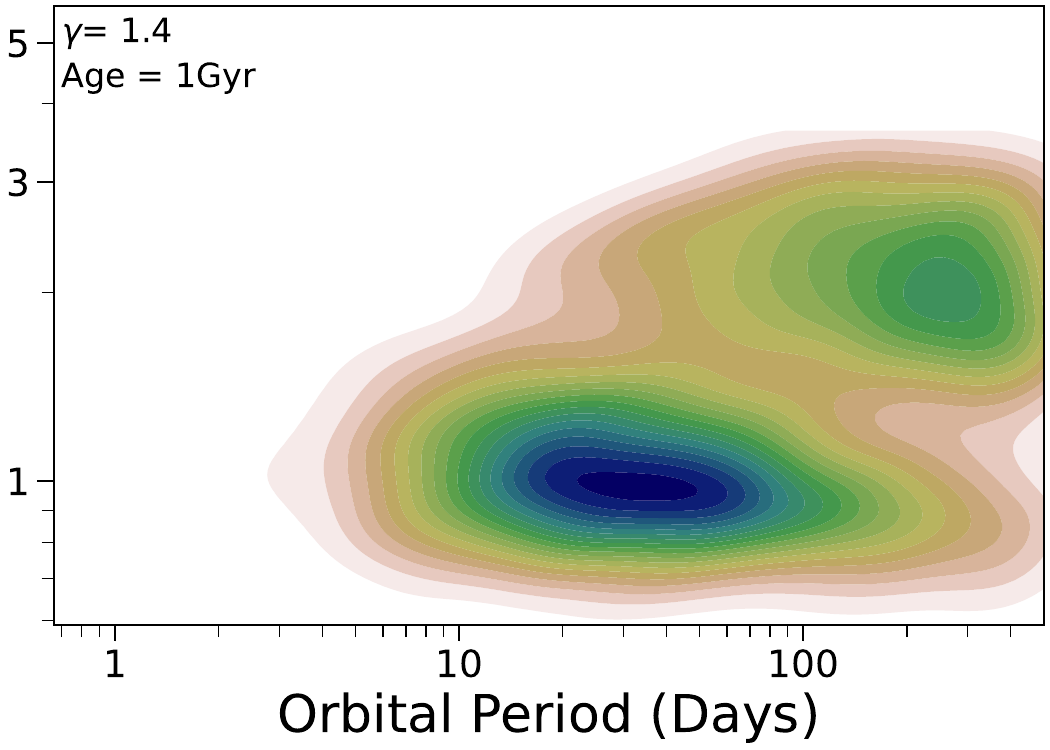}{0.5\textwidth}{}}
\caption{Same as Figure \ref{fig:Radius_kde_5800K} but with photoevaporation. A dramatic decrease in sub-Neptunes and the corresponding increase in super-Earths is observed for $\gamma = 1.4$, filling in the radius valley even more, especially at short orbital periods. The increase in super-Earths is also pronounced for $\gamma = 1.2$ but the long period ($\gtrsim 100$ days) sub-Neptunes are more preserved.}
\label{fig:Photoevaporation_kde}
\end{figure*}

\begin{figure*}
\epsscale{1.1}
\plottwo{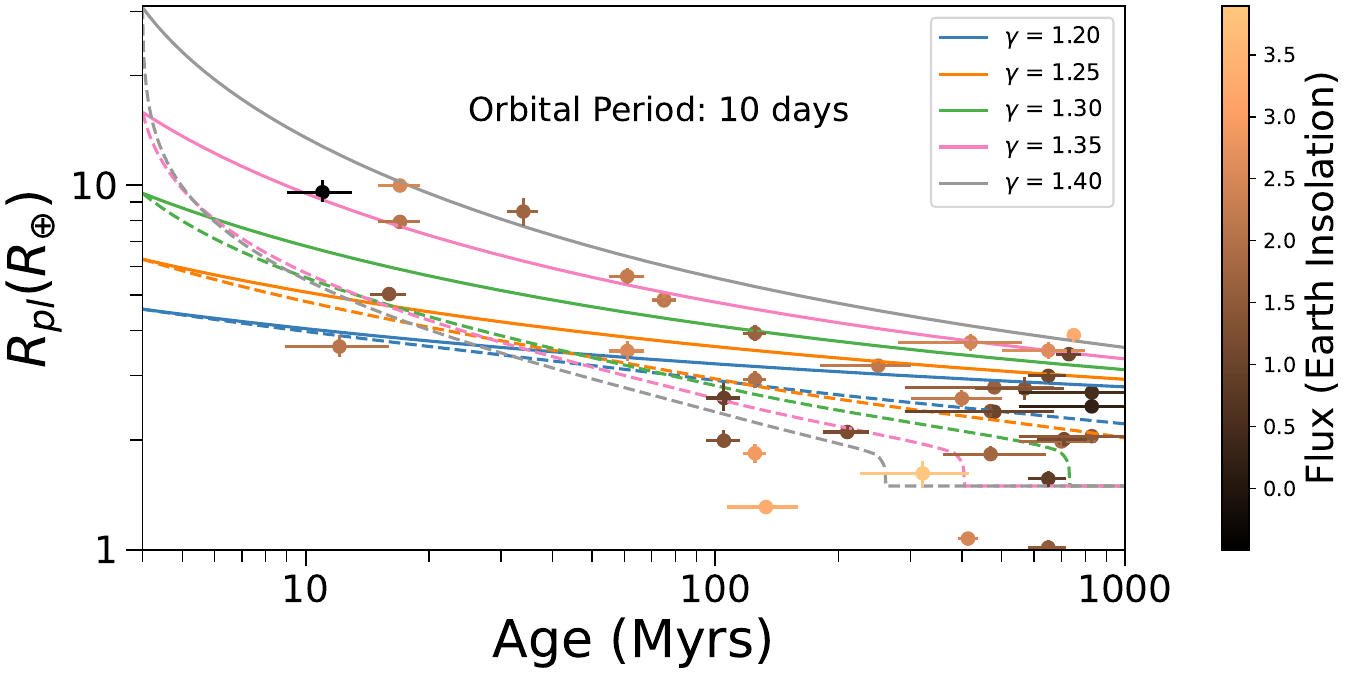}{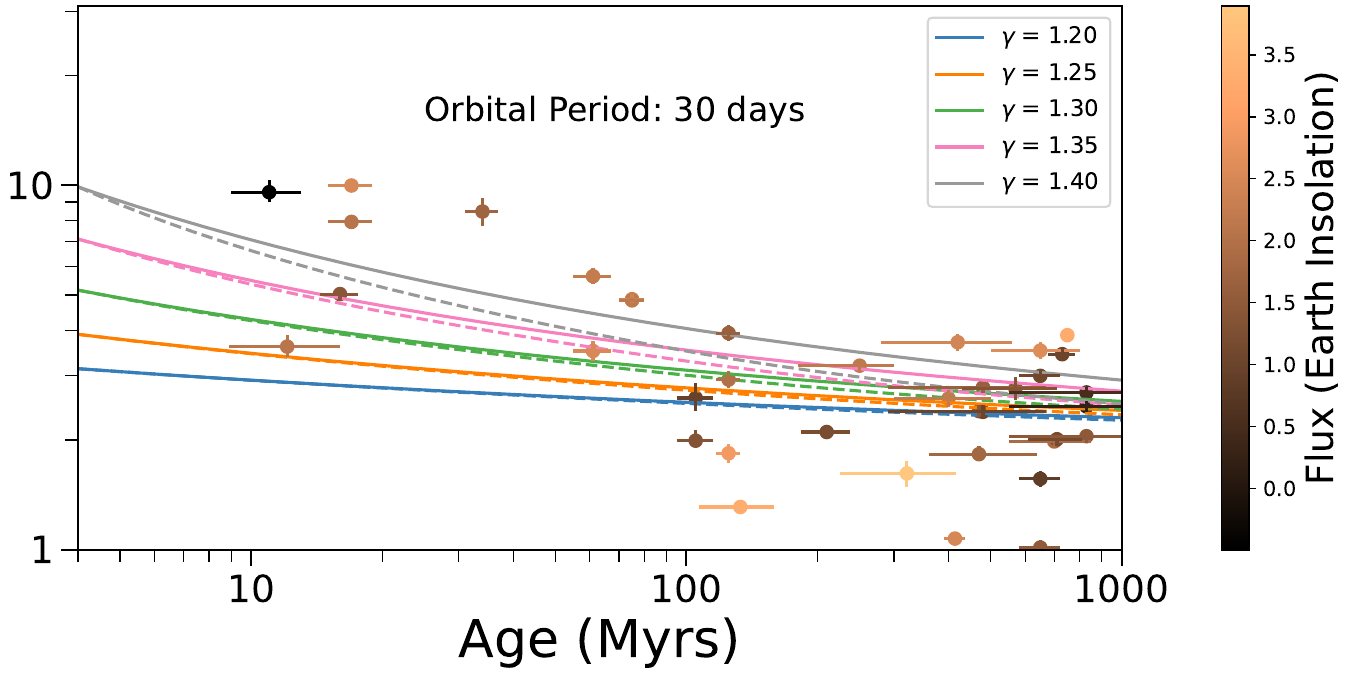}
\
\caption{Our time evolution model with and without mass loss (photoevaporation only) compared with data from exoplanets in the NASA exoplanet archive retrieved on 2026-05-28 \citep{NASA}. The solid lines correspond to the expected radius evolution under passive cooling without mass loss and the dashed lines correspond to that from photoevaporative mass loss. For illustration purpose, we fix the planet properties to $M_{\rm core}=5M_{\oplus}$ and $M_{\rm env}/M_{\rm core}=5\%$ in our calculations at 10 days (left) and at 30 days (right). Different colored lines correspond to different $\gamma$'s while the datapoints are color-coded with respect to the incident flux normalized to that of the Earth.}
    \label{fig:data}
\end{figure*}

\section{Discussion and Conclusion} \label{sec:Discussion}

In this paper, we have shown how the radius evolution of a planet is dependent on the adiabatic index of its envelope which probes the internal mass distribution. An adiabatic index $\gamma < 4/3$ corresponds to a centrally concentrated mass profile, leading to shallower radius relations in regards to both time evolution and the total envelope mass. As a consequence, although planets of higher $\gamma$ initially start larger, those with very thin atmosphere (envelope mass fraction $M_{\rm env}/M_{\rm core} \lesssim$10$^{-3}$) appear more compact for higher $\gamma$ over $\sim$10--100 Myr timescale. For thicker atmospheres, planets of different $\gamma$ converge to a similar radius over $\sim$1 Gyr.

Since the rate of mass loss is larger for larger planets, those with high $\gamma$ lose their envelope more quickly and can quickly transform into rocky planets at short orbital periods $\sim$10 days, especially under photoevaporation. At longer periods, the convergence of radius between envelopes of different $\gamma$ is achieved more quickly compared to passive cooling (i.e., no mass loss).

We close our paper with implications of our results on the observables such as the radius valley and the radius of young planets.

\subsection{The Radius Valley}

The observed valley in the radius-period space is understood as the division between smaller rocky planets and the larger gas-enveloped sub-Neptunes \citep{Fulton2017,Fulton2018,vaneylen2018}. Given the difference in planet radius for different adiabatic indices, we explore how the adoption of different $\gamma$ changes the shape of the radius valley.

We construct a population of 5000 planets using the methodology outlined in \citet{Lee2022} which computes the expected envelope mass fraction following nebular gas accretion for a given disk profile, planet core mass, and orbital period. Core masses are drawn from the distribution
\begin{equation}
    \frac{dN}{d\log M_{\rm core}} \propto M_{\rm core}^{\delta} \exp\left(-\frac{M_{\rm core}}{M_{\rm break}}\right)
    \label{eq:dNdlogMc}
\end{equation}
with $\delta=0.3$ and $M_{\rm break}=4M_\oplus$. The overall functional form is chosen to describe the fall-off in the number of $\gtrsim$10--20$M_\oplus$ sub-Neptunes while allowing for the possibility of bottom-heavy mass function (see \citealt{Lee2022} for more detail). 

We fix the host stellar mass to solar mass and set the minimum core mass to 0.5$M_\oplus$. Stellar luminosity evolution follows that from \citet{Johnstone21}. Cores begin accreting gas when the disk gas is depleted from the minimum mass extrasolar nebula by 0.0001 ($f_{\rm dep,1}$ in \citet{Lee2022}). Planet orbital periods $P$ are drawn from the following distribution from 0.5 to 500 days:
\begin{equation}
    \frac{dN}{d\log P} \propto 1-\exp\left[-\left(\frac{P}{11.9\,{\rm days}}\right)^{2.4}\right],
    \label{eq:dNdlogP}
\end{equation}
which modifies the fitting to the planet occurrence rate reported by \citet{Petigura18} to enforce a log-uniform distribution of $P$ beyond $\sim$10 days. Of the 5000 generated planets, we only consider those whose initial envelope mass fraction lies below 10\% as our analytic calculation of radius, which ignores envelope self-gravity, overestimates the radius significantly at higher values. We end up with 4324 planets whose radius we evolve at varying $\gamma$ following our methods described in Section \ref{sec:Methods}.

Figure \ref{fig:Radius_kde_5800K} presents the radius-period distribution at 100 Myr and 1 Gyr under passive cooling (i.e., no mass loss) for our minimum and maximum $\gamma$. In general, we see no difference in either the location or the slope of the radius valley between different $\gamma$'s, which is expected given that the initial radius valley by gas accretion physics occurs at envelope mass fraction $\sim$10$^{-4}$--10$^{-3}$ \citep[see][their Figure 9]{Lee2022} at which point the atmosphere is so thin that the effect of $\gamma$ on planet radius is muted (see Figure \ref{fig:Radius_env}).

Where the two $\gamma$'s differ is in the depth of the valley where minute differences in planet radii can matter. We find that the radius valley appears to close up for lower adiabatic indices at younger ages ($\sim$100 Myr), but then begins to close up more for higher adiabatic indices at older ages ($\sim$1 Gyr). Such a behavior can be understood as an extension of our results in Figure \ref{fig:radius_evo}. Since the planets at the inner edges of the valley are those of small envelope mass fraction, they will appear slightly smaller with lower $\gamma$ at younger ages but the more rapid contraction of high $\gamma$ envelopes means these edge planets will appear smaller for higher $\gamma$ at older ages, effectively filling in the radius valley at a more evolved stage.

The result of mass loss by photoevaporation applied to the same population of planets is shown in Figure \ref{fig:Photoevaporation_kde}. We see a clear signal of the more dramatic transformation of short-period ($\lesssim$20--30 days) sub-Neptunes into rocky planets for higher $\gamma$ at both 100 Myr and 1 Gyr, owing to the more rapid mass loss for planets of larger adiabatic indices (Section \ref{ssec:impact-mloss}). This $\gamma$-dependent mass loss rate manifests as different shapes of the radius valley between low and high $\gamma$. In case of $\gamma=1.2$, while we see a significant increase in the population of super-Earths and rocky planets and clearing of the valley compared to the no-mass-loss scenario (Figure \ref{fig:Radius_kde_5800K}), the slope and the location of the valley is largely maintained. In case of $\gamma=1.4$, by contrast, we see a closing of the valley at $\sim$20--30 days and a widening of the valley at longer orbital periods, effectively straightening the inner edge of the sub-Neptune side of the radius valley. Current exoplanet data for small planets are limited in sensitivity at periods beyond $\sim$50 days so it is not clear if the width of the radius valley is period-variant. Increasing the sample of data at these long periods (by e.g., Transiting Exoplanet Survey Satellite (TESS), PLAnetary Transits and Oscillations of stars (PLATO)) would be a welcome endeavor to constrain the initial planet properties.

\subsection{Young Planets}

The effect of different $\gamma$ on planet radius is most pronounced at young ages. We compare our simple analytic model against the confirmed exoplanets around young stars in Figure \ref{fig:data}. These exoplanets were retrieved from the \citet{NASA} with the following conditions: orbital periods shorter than 300 days, measurements of planet radii, mass less than 0.1 Jupiter masses if measurements exist, measurements of stellar effective temperature and age that is less than 1 Gyr. We take the default parameters provided by the archive. We additionally remove Mascara-1 b ($R_{\rm pl} \sim$17$R_\oplus$, 800 Myr) which has a measured mass of $\sim$3.7$M_{\rm Jup}$ \citep{Talens17}.

While we see the overall decrease in the size of the planets as a function of time that closely follows the model expectation from passive cooling, the scatter in data spans the full range of $\gamma$ we consider. The increased population of $\lesssim$2$R_\oplus$ planets beyond 100 Myr broadly agrees with the expectation of photoevaporative mass loss with the caveat that the lack of such planets at younger ages may be an observational bias.\footnote{We show only the photoevaporative model and not the bolometric loss model because, for an orbital period of 10 days, our bolometric loss model is unable to completely shed the atmosphere for our chosen planet parameters---clearly in contradiction with the observations. In addition, we observe a negligible difference between the photoevaporative and bolometric loss models at an orbital period of 30 days.}  

Ultimately, we expect the adiabatic index to change throughout planet's thermal evolution. Thermodynamic models of initial gas accretion find the initial interior $\gamma \sim 1.2$ when the planet's interior temperature is much higher causing H$_2$ dissociation \citep{Lee14,Lee15}. As the envelope cools, we expect gas to become more molecular which would increase $\gamma$. The significant difference in the planet radius with respect to time and envelope mass fraction for different adiabatic indices that we find suggests that the planet's initial thermal state (particularly as it pertains to internal mass distribution) is an important factor to consider in studies that attempt to use exoplanet radius measurements to trace back their formation conditions and mass evolution, especially as the sample of observed young planets grows with e.g., TESS. Future studies that self-consistently couple the calculation of gas envelope accretion with subsequent thermal evolution with an explicit investigation of the changes in the interior structure profile is suggested to address this issue.

\begin{acknowledgments}
We thank the anonymous referee whose report helped improve the manuscript.
AC thanks Regent Scholars for their support in enabling this work through the Regents Scholars Research Initiative at UC San Diego. EJL was supported by NSF Research Grant 2509275.
This research was enabled in part by support provided by the BC DRI Group and the Digital Research Alliance of Canada (alliancecan.ca) through the fir cluster. This research has made use of the NASA Exoplanet Archive, which is operated by the California Institute of Technology, under contract with the National Aeronautics and Space Administration under the Exoplanet Exploration Program.
\end{acknowledgments}

\begin{contribution}
AC carried out all the analytic and semi-analytic calculations and drafted much of the paper. EJL conceived the project, provided the model population of planets, and supervised the analysis and writing.
\end{contribution}

\software{astropy \citep{2013A&A...558A..33A, 2018AJ....156..123A},
          Matplotlib \citep{Hunter2007},
          Numpy \citep{harris2020array}
          Scipy\citep{Scipy}
          Pandas \citep{Pandas}}

\bibliography{rvalley-gamma}{}
\bibliographystyle{aasjournalv7}

\end{document}